\documentstyle[psfig]{article}

\newcommand{\tr} {\mbox{${\rm Tr}$}}
\newcommand{\bp} {\mbox{\boldmath$p$}}
\newcommand{\bq} {\mbox{\boldmath$q$}}
\newcommand{\bx} {\mbox{\boldmath$x$}}
\newcommand{\bxi}{\mbox{\boldmath$\xi$}}
\newcommand{\Det}{\mbox{${\rm Det}\:$}}
\newcommand{\Dslash} {\mbox{$D\hskip -0.65em /\;$}}

\title{Geometrical properties of Maslov indices in periodic-orbit theory}
\author{Ayumu Sugita\thanks{sugita@ruby.scphys.kyoto-u.ac.jp}\\
Department of Physics,Graduate School of Science,Kyoto University,\\
Kyoto 606-01,Japan}

\begin{document}

\maketitle

\begin{abstract}
Maslov indices in periodic-orbit theory  are 
investigated using phase 
space path integral. Based on the observation that the Maslov index is 
the multi-valued function
of the monodromy matrix, we introduce 
a generalized monodromy matrix in 
the universal covering space of the symplectic group and show that
this index is uniquely determined in this space. 
The stability of the orbit is shown to determine the parity of the index,
and a formula for the index of the n-repetition of the orbit is derived.
\end{abstract}

\vspace{0.5cm}
\hspace{1.0cm} PACS numbers: 05.45.Mt, 02.40.Ma 
\vspace{0.5cm}

 Over the past few decades a large number of studies have been made
on periodic-orbit theory \cite{gutzwiller}. However, the  Maslov index, 
which is an additional phase factor appearing in periodic orbit theory,
doesn't seem to be thoroughly understood. 
For hyperbolic orbits, Robbins \cite{robbins} showed that these
indices are 
the winding numbers which are defined by 
the invariant Lagrangian manifolds
around these orbits. Moreover, it was conjectured that the same argument
could be extended to elliptic orbits and more general orbits which have
mixed stability. 
However, this is not the case
since general orbits don't necessarily have such invariant manifolds 
around them. 
Furthermore, Brack et al. \cite{brack} investigated periodic orbits
in anisotropic harmonic oscillator (these orbits are elliptic), and 
showed the Maslov index of the n-repetition of the orbit
$\mu_{n}$ is not equal to $n\mu_{1}$.
This result  contradicts Robbins's conjecture, which
leads to $\mu_{n} = n\mu_{1}$.

In this letter, we propose a new approach to the problem
which can be applied to
all periodic orbits, irrespective of the type of the stability.
Our method is based on phase space path integral of the partition
function. The advantage of using phase space path integral is that
it respects symmetries of the system: 
time translation symmetry and canonical invariance.
In addition
to global canonical transformations, our method allows us to use
local canonical transformations which act differently at each point of
the orbit. They are regarded as gauge transformations with gauge group 
${\rm Sp}(2n,R)$, and the topological nature of this gauge group 
plays essential role in our theory. 
Derivation of semiclassical trace formula using 
phase space path integral was 
given by Sakita et al.\cite{sakita}, 
and similar analysis was made by Kuratsuji 
\cite{kuratsuji} for coherent state path integral, but Maslov
indices are neglected in these works. 
Maslov indices of semicalassical
propagator (but not the periodic orbit)
was investigated by Levit et al. \cite{levit1,levit2}
using phase space path integral, and they showed that this index
is given by the excess of the negative eigenvalues of 
second variation of action around the orbit.
We treat Maslov indices of periodic orbits 
in the same way as theirs. There are two types of Maslov
indices of periodic orbits: indices for the partition function
(time region), and that for the density of states (energy region).
We calculate Maslov indices for the partition function first.
Then it is easy to calculate Maslov indices for the density of
states by taking into account the extra phase contributions that
arises in performing the Fourier-Laplace transform by the
stationary phase approximation.

The main idea of this paper is that the linearized symplectic
flow around the orbit directly
determines the Maslov index. Therefore we don't use
Lagrangian manifolds in this paper. 
The linearized symplectic flow
is represented by $V(t)$ (a symplectic matrix with one parameter $t$),
and $V(t)$ is considered to be a curve on the group manifold of 
${\rm Sp(2n,R)}$. We can see various properties of Maslov indices
by continuous deformation of this curve.

Our final results can be summarized in the following formula 
eq.(\ref{1}) and eq.(\ref{2}).
The Maslov index of the orbit is 
\begin{equation}
\mu = p + q + 2k,
\label{1}
\end{equation}
where $p$ is the numbers of elliptic pairs of eigenvalues, 
$q$ is the number of 
inverse hyperbolic pairs of eigenvalues, and $k$ is the 
winding number which will
be defined later on. 
We can see that parity of the index is determined
by $p$ and $q$ from this formula. 
The Maslov index of the n-repetition of this orbit is
\begin{equation}
\mu_{n} = \sum_{i=1}^{p}
\left(1 + 2\left[\frac{n\alpha_{i}}{2\pi}\right]\right)
+ nq + 2nk,
\label{2}
\end{equation}
where $\alpha_{i}$ is the stability angle of the i-th elliptic eigenvalues.
Note that $\mu_{n}$ is not equal to $n\mu_{1}$ if the orbit have elliptic 
eigenvalues.

Let us start with the following identity for the density of states;
\begin{equation}
\rho (E) = - \frac{1}{\pi} {\rm Im} \; g(E+i\epsilon),
\end{equation}
\begin{equation}
g(E) = \tr \frac{1}{E - \hat{H}}.
\end{equation}
$g(E)$ is obtained by Fourier-Laplace transform of the partition function:
\begin{equation}
g(E) = \frac{1}{i\hbar}\int_{0}^{\infty}dT e^{iET/\hbar} Z(T).
\label{fourier}
\end{equation}
We use the phase space path integral of the partition function:
\begin{equation}
Z(T) = {\rm Tr} \exp \left[- \frac{i}{\hbar}\hat{H}T \right] = 
\int {\cal D}\bp {\cal D}\bq \exp 
\left[ \frac{i}{\hbar}\oint (\bp d\bq - H dt)\right].
\label{path_int}
\end{equation}
Strictly speaking, we should write (\ref{path_int}) 
in a discrete form to obtain 
well-defined continuum limit.
However, to make the discussion simple and show our essential idea,
we use continuum notation throughout this paper.
The detailed analysis of (\ref{path_int}), 
including subtle problems concerning 
operator ordering and changing variables, 
will be reported in the forthcoming paper. \cite{sugita}.

\begin{figure}
\hskip 3cm 
\psfig{file=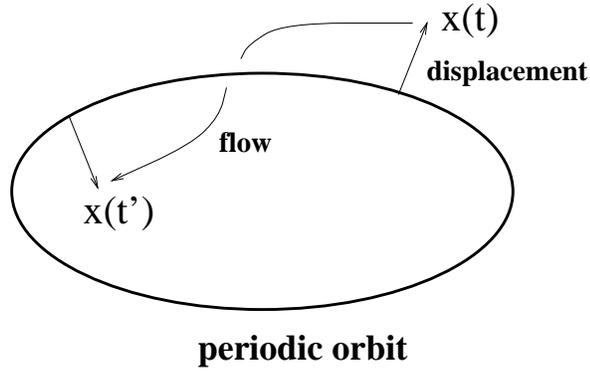,height=0.4\textwidth,angle=-90}
\caption{The space of the displacement vectors is regarded as fibre
bundle, and Hamiltonian symplectic flow defines the connection. The
structure group of this space is ${\rm Sp}(2n,R).$}
\label{bundle}
\end{figure}

Let us evaluate (\ref{path_int}) by
the stationary phase approximation. The stationary phase condition reads
\begin{equation}
\delta \oint (\bp d\bq - H dt) = 0,
\label{variation}
\end{equation}
which leads to the Hamilton equations of motion. 
It follows from the periodic boundary condition 
of the path integral that the 
solutions of (\ref{variation}) are periodic orbits.
Thus the semi-classical approximation $Z_{sc}(T)$ to $Z(T)$ is given 
by a sum over the periodic orbits.
\begin{equation}
Z_{sc}(T) = \sum_{p.o.} K  \exp \left[\frac{i}{\hbar} R \right].
\end{equation}
Here $R = \oint (\bp d\bq - H dt)$ is the classical action of the 
periodic-orbit, and $K$ is the contribution of the quadratic 
fluctuation around the orbit,
\begin{equation}
K = \int {\cal D}\bx \exp\left[i\delta^{2} R[\bx(t)] \right],
\label{path_int2}
\end{equation}
\begin{equation} 
\bx (t) = \frac{1}{\sqrt{\hbar}}
(\delta p_{1},\delta q_{1},\delta p_{2},.....,\delta q_{n})
\end{equation}
The explicit form of the second variation $\delta^{2}R$ is
\begin{equation}
\delta^{2}R[\bx (t)] = \frac{1}{2}\int_{0}^{T}dt\; 
\bx^{T}(t)\left(-J\frac{d}{dt} - H^{''}(\bp(t),\bq(t))\right)\bx(t),
\label{delta2R}
\end{equation}
where $J$ is a matrix of the form:
\begin{equation}
J = \left(
\begin{array}{ccccc}
j & & & & \\
& j & & & \\
& & . & & \\
& & & . & \\
& & & & j 
\end{array}
\right),
\end{equation}
and $j$ is a $2\times 2$ matrix:
\begin{equation}
j = \left(
\begin{array}{cc}
0 & -1 \\
1 & 0
\end{array}
\right).
\end{equation}
We rewrite (\ref{delta2R}) as 
\begin{equation}
\delta^{2}R = 
-\frac{1}{2}\int_{0}^{1}dt^{'} \bx^{T}(t^{'})\Dslash (t^{'})\bx(t^{'}),
\end{equation}
where $t^{'}=t/T$, and $\Dslash$ is defined as
\begin{equation}
\Dslash(t) = JD = J\left(\frac{d}{dt} - A(t)\right),
\label{Dsla} 
\end{equation}
\begin{equation}
A(t) = T J \; H^{''}(\bp(t/T),\bq(t/T)).
\label{A1}
\end{equation}
$A$ is the connection for the displacement vector $\bx$, and $D=d/dt - A$ is
the covariant derivative.
The parallel transport of $\bx$ is defined by the equation $D\bx=0$  
(Fig.\ref{bundle}),
which is the Hamiltonian equation of motion for $\bx$:
\begin{equation}
\frac{d}{dt} \bx = J H^{''}\left(\bp(t),\bq(t)\right) \bx .
\end{equation}  
Hereafter we slightly generalize the problem, and $A$ is
considered to be a generator of symplectic group 
which satisfies $A^{T}J + JA = 0$. Then $A$ is written as  
\begin{equation}
A(t) = JH(t),
\label{A2}
\end{equation}
where $H$ is a symmetric matrix of even demension.
(\ref{A1}) is the special case of (\ref{A2}). 
We deal with the quadratic path 
integral of the following form instead of (\ref{path_int2}):
\begin{equation}
K = \int {\cal D}\bx \exp \left[-\frac{i}{2}\bx^{T}\Dslash \bx\right] = 
\frac{1}{\sqrt{\Det i\Dslash}} 
\label{K}
\end{equation}
Here $\Dslash = J(d/dt - A)$, and $A$ is defined in (\ref{A2}).
The phase of
(\ref{K}) defines the Maslov index:
\begin{equation}
\frac{1}{\sqrt{\Det i\Dslash}}  = 
\frac{e^{-i\pi\mu /2}}{\sqrt{|\Det \Dslash|}}
\label{main}
\end{equation}
\begin{equation}
\mu = \frac{\mu_{+}-\mu_{-}}{2}.
\label{def}
\end{equation}
Here $\mu_{+}(\mu_{-})$ denote the number of the positive (negative)
eigenvalues of $\Dslash$.
In the continuum limit, both $\mu_{+}$ and 
$\mu_{-}$ become infinity, and the formula (\ref{def}) 
cannot directly define the 
Maslov index. However, we can evaluate the relative change of the index
by deforming the operator $\Dslash$ continuously and analyzing the
points where the sign of the eigenvalue changes. 
(This is similar to the discussion given in \cite{levit1,levit2} 
for the semi-classical propagator.)

In the following, we calculate $|\Det\Dslash|$ and $\mu$ in (\ref{main}).
The procedure consists of three steps.
\begin{enumerate}
\item Diagonalization of  $D$ and calculation of  $|\Det\Dslash| = |\Det D|$.
\item Diagonalization of  $\Dslash$ using the basis given in the first step.
\item Calculation of $\mu$ by continuous deformations of $\Dslash$.
\end{enumerate}
The first step is essentially the same as the discussions given in 
\cite{sakita}
and \cite{kuratsuji}. The second and the third steps are first
given in this paper.

The first step is the calculation of $|\Det\Dslash|$.
For this purpose, we first diagonalize 
the covariant derivative $D$:
\begin{equation}
D\bx = \left(\frac{d}{dt} - A\right)\bx = \epsilon \bx.
\label{ddiag}
\end{equation}
This equation is easy to solve because the time derivative has diagonal form.
We introduce a symplectic matrix $V(t)$ which
stands for the flow around the orbit:
\begin{equation}
D V(t) = 0.
\end{equation}
If we fix $V(0)=1$, $D$ and $V$ has one to one correspondence,
and $V(1) \;(= M )$ is the monodromy matrix of the orbit.
Let $\bxi_{\zeta}$ be the eigenvector of M, and $\zeta $ be 
the stability angle 
of the orbit,
\begin{equation}
M \bxi_{\zeta} = e^{-i\zeta} \bxi_{\zeta}.
\end{equation}
Then the solutions of (\ref{ddiag}) are written as follows:
\begin{equation}
\bx_{\zeta,n}(t) = \exp (\epsilon_{\zeta,n} t) V(t) \bxi_{\zeta},
\label{base}
\end{equation}
\begin{equation}
D \bx_{\zeta,n} = \epsilon_{\zeta,n} \bx_{\zeta,n},
\label{eigen1}
\end{equation}
\begin{equation}
\epsilon_{\zeta,n} = ( \zeta + 2n\pi) i.
\label{eigenvalue}
\end{equation}
If $e^{-i\zeta}$ is the eigenvalue of $M$, $e^{i\zeta}$ also becomes
the eigenvalue of $M$ \cite{arnold}, and the absolute value of the 
functional determinant of $\Dslash$ can be evaluated as a product
of eigenvalues (\ref{eigenvalue}):
\begin{eqnarray}
|\Det\Dslash | = |\Det D| & = & \prod_{n,l}
\left|(2\pi n)^{2} - \zeta_{l}^{2}\right|, \\
& \sim & 
\left| \prod_{l}\zeta_{l}^{2}\prod_{n=1}^{\infty}
\left(1 - \left(\frac{\zeta_{l}}{2n\pi}\right)^{2} \right)\right|. 
\end{eqnarray}
Here we replace the divergent constant 
$\prod_{n=1}^{\infty} 2\pi n$ by 1. 
In discrete formalism,
this term corresponds to 
\begin{equation}
\frac{1}{N}
\prod_{n=1}^{N-1} 2\sin \frac{\pi n}{N}
=1.
\label{renorm}
\end{equation}
For $n\ll N$, $\prod_{n} 2\sin\pi n/N \sim \prod_{n} 2\pi n/N$, 
therefore 
(\ref{renorm}) is considered to be a normalized value of
$\prod_{n=1}^{\infty} 2\pi n$.
\footnote{This rule
can also be derived from analytic continuation of a generalized zeta function.
Let $\zeta_{a}(s)=\sum a_{n}^{-s}$, then  $\exp(-\zeta_{a}^{'}(0))$ 
is regarded
as a normalized value of $\prod_{n} a_{n}$. 
In this case, $a_{n}=2\pi n$
and $\zeta_{a}(s) = (2\pi)^{-s}\zeta (s)$.
($\zeta$ is Riemann zeta function.)
Therefore 
$\zeta_{a}^{'}(s)=(2\pi)^{-s}(\zeta^{'}(s) - \zeta (s)\ln (2\pi))$ and
$\zeta_{a}^{'}(0) = 0$.(Here we use $\zeta (0)= -1/2$ and 
$\zeta^{'}(0)=-\ln\sqrt{2\pi}$.) 
After all we obtain $\exp(-\zeta_{a}^{'}(0))=1$.}
Details of this renormalization procedure will be given in \cite{sugita}.

The result is
\begin{eqnarray}
|\Det\Dslash | & = & 
\left|\prod_{l}\left(2\sin \frac{\zeta_{l}}{2}\right)^{2}\right|, \\
& = & |\det (M - I)|.
\end{eqnarray}
Here we used the formula 
\begin{equation}
\prod_{n=1}^{\infty}\left(1-\frac{x^{2}}{n^{2}}\right)
= \frac{\sin\pi x}{\pi x}.
\end{equation}

The second step is the diagonalization of $\Dslash$. 
If we represent $\Dslash$ using the basis (\ref{base}), 
it is easy to see that $\Dslash$ is the block diagonal matrix.
\footnote{In general, a real symplectic vector space with a given quadratic 
form can be decomposed into a direct sum of skew orthogonal real symplectic
subspaces, and the quadratic form is represented as a sum of normal
forms on these subspaces. This is known as Williamson's theorem 
\cite{arnold,williamson}.}
We can obtain diagonal representation of $\Dslash$ by diagonalizing
this block diagonal matrix again.

First we assume that $M$ has no degenerate eigenvalue.
If $\zeta$ is a stability angle of the orbit, $-\zeta$ and $\zeta^{*}$
are also stability angles of the orbit because $M$ is the symplectic matrix
\cite{arnold}. Therefore eigenvalues of $M$ can be classified into four types,
and $M$ can be transformed into a block diagonal matrix by a symplectic
transformation.
\begin{equation}
M = \left(
\begin{array}{ccccc}
m_{1} & & & & \\
& m_{2} & & & \\
& & .& & \\
& & & .& \\
& & & & m_{k} \\
\end{array}
\right)
\end{equation}
\begin{enumerate}
\item elliptic: $\zeta = \pm \alpha$
\begin{equation}
m = \left(
\begin{array}{cc}
\cos\alpha & -\sin\alpha \\
\sin\alpha & \cos\alpha
\end{array}
\right)
\end{equation}
\item hyperbolic: $\zeta = \pm i\beta$
\begin{equation}
m = \left(
\begin{array}{cc}
e^{\beta} & 0 \\
0 & e^{-\beta}
\end{array}
\right)
\end{equation}
\item inverse hyperbolic: $\zeta = \pm i(\beta + \pi)$
\begin{equation}
m = \left(
\begin{array}{cc}
- e^{\beta} & 0 \\
0 & - e^{-\beta}
\end{array}
\right)
\end{equation}
\item loxodromic: $\zeta = \pm \alpha \pm i\beta$ 
\begin{equation}
m = \left(
\begin{array}{cccc}
e^{\beta}\cos\alpha & 0 & -e^{\beta}\sin\alpha & 0 \\
0 & e^{-\beta}\cos\alpha & 0 & -e^{-\beta}\sin\alpha \\
e^{\beta}\sin\alpha & 0 & e^{\beta}\cos\alpha & 0 \\
0 & e^{-\beta}\sin\alpha & 0 & e^{-\beta}\cos\alpha 
\end{array}
\right)
\end{equation}
\end{enumerate}
The matrix elements between $\bx_{\zeta,n}$ and
$\bx_{\zeta^{'},n^{'}}$ are non-zero 
only if $\zeta$ and $\zeta^{'}$
belong to the same block. Therefore it is enough to diagonalize each block
separately to diagonalize $\Dslash$. 

For example, eigenvalues and eigenvectors which belong to
a hyperbolic block are
\begin{equation}
M\bx_{\pm} = e^{\pm\beta}\bx_{\pm}.
\end{equation}
Here, the normalization condition is taken as
\begin{equation}
[\bx_{+},\bx_{-}] = 1,
\end{equation}
and eigenvalues and eigenvectors of $D$ are
\begin{equation}
D \bx_{\pm,n}(t) = \epsilon_{\pm,n} \bx_{\pm,n}(t),
\end{equation}
\begin{equation}
\bx_{\pm,n}(t) = \exp (\epsilon_{\pm,n})V(t)\bx_{\pm}(t),
\end{equation}
\begin{equation}
\epsilon_{\pm,n} = \mp\beta + 2n\pi i.
\end{equation}
We define real vectors $\bx_{c,\pm,n},\bx_{s,\pm,n}$ as 
\begin{eqnarray}
\bx_{c,\pm,n} & = & \frac{1}{\sqrt{2}}(\bx_{\pm,n} + \bx_{\pm,-n}), \\
\bx_{s,\pm,n} & = & \frac{1}{\sqrt{2}i}(\bx_{\pm,n} - \bx_{\pm,-n}) \;\;\; 
(n=1,2,...). 
\label{realbase}
\end{eqnarray}
Since $\bx_{\pm,0}$ are real vectors, we don't have to
re-define them. 
Matrix elements which have non-zero values are
\begin{eqnarray}
\int_{0}^{1}\! dt \; \bx_{c,+,n}^{T} \Dslash \bx_{c,-,n} 
& = & - \beta,\\
\int_{0}^{1}\! dt \; \bx_{s,+,n}^{T} \Dslash \bx_{c,-,n}
& = & - 2n\pi,\\
\int_{0}^{1}\! dt \; \bx_{c,+,n}^{T} \Dslash \bx_{s,-,n}
& = & 2n\pi,\\
\int_{0}^{1}\! dt \; \bx_{s,+,n}^{T} \Dslash \bx_{c,-,n}
& = & - \beta, \\
\int_{0}^{1}\! dt \; \bx_{+,0}^{T} \Dslash \bx_{-,0}
& = & - \beta .
\end{eqnarray}
Therefore $\Dslash$ is represented by the matrix $d_{n}$ 
in the space spanned by 
$\bx_{c,\pm,n},\bx_{s,\pm,n}$: 
\begin{equation}
d_{n} = 
\left(
\begin{array}{cccc}
0 & 0 & - \beta & 2n\pi\\
0 & 0 & - 2n\pi & - \beta\\
- \beta &  - 2n\pi & 0 & 0 \\
2n\pi & - \beta & 0 & 0
\end{array}
\right)\;\;\;(n\ge 1).
\end{equation}
In the case of $n=0$,
\begin{equation}
d_{0} =
\left(
\begin{array}{cc}
0 & - \beta \\
- \beta & 0 \\
\end{array}
\right) .
\end{equation}
The solutions of the eigenvalue equation $\det (d_{n}-\lambda I)=0$
are
\begin{eqnarray}
\lambda & = & \pm \sqrt{\beta^{2} + (2n\pi)^{2}} \;\;\;\;(n\ge 1), \\
\lambda & = & \pm \beta \;\;\;\; (n=0) .
\end{eqnarray}
The solutions for $n\ge 1$ are doubly degenarate.
Thus we obtain diagonal representation of $\Dslash$ for hyperbolic
blocks.

Other blocks can be treated in the same way. First, we define
real basis from the real parts and the imaginary parts of the
basis ($\ref{base}$), and represent $\Dslash$ using this basis.
Then $\Dslash$ become a block diagonal matrix and each block
is $4\times 4$ at most. Second, we diagonalize these blocks, 
then we obtain diagonal representation of $\Dslash$.

Diagonal elements of an elliptic block are
\begin{equation}
\alpha + 2n\pi \;\;\; (n=0,\pm 1,\pm 2,...).
\end{equation}
For a inverse hyperbolic block, 
\begin{equation}
\pm \sqrt{\beta^{2} + (2n+1)^{2}} \;\;\; (n=0,1,2,...).
\end{equation}
For a loxodromic block,
\begin{equation}
\pm \sqrt{\beta^{2} + (\alpha+2n\pi)^{2}} \;\;\; (n=0,\pm 1,\pm 2 ...).
\end{equation}
All diagonal elements are doubly degenerate. These results are
summarized in Table \ref{table}.
\footnote{Note that the absolute values of the diagonal elements has 
no generic meaning because we use the non-orthogonal 
(symplectic) basis. On the other hand, the signs of the diagonal elements have
generic meaning.}

Till now, we have assumed that the monodromy matrix $M$ has
no degenerate eigenvalue.
If $M$ has degenerate eigenvalues, we have to deal with 
these cases separately. We will not go into the details of these
exceptional cases, but
the case in which $M$ has eigenvalue 1 is very important.
\footnote{The eigenvalue 1 is always degenerate; The number
of other eigenvalues ($\ne 1$) is even, therefore the number of
the eigenvalues equal to 1 is even.}
In this case, the stability angle of the eigenvalue is 0,
which means that $\Dslash$ USA a zero-mode and the orbit is
not isolated. 
Therefore we have to calculate the functional
determinant of $\Dslash$ in the space where the zero-mode is removed
(we denote this as $\Det^{'}\Dslash$), and integrate with respect
to the zero-mode separately. 

We deal with the simplest case in which the eigenvalue 1 is doubly
degenerate.(This type is called parabolic.) 
We can choose the basis $\bx_{\alpha}\;(\alpha = 1,2)$ as
\begin{eqnarray}
M \bx_{1} & = & \bx_{1}, \\
M \bx_{2} & = & \bx_{2} + \gamma \bx_{1},
\end{eqnarray}
\begin{equation}
[\bx_{1},\bx_{2}] = 1.
\end{equation}
In this case, the basis (\ref{base}) are not complete set.
Therefore we define new basis as
\begin{eqnarray}
\bx_{\alpha,n}(t) & = & 
V(t)\exp \left[t(2n\pi i I - B)\right]\bx_{\alpha} 
\;\;\;(\alpha = 1,2).
\end{eqnarray}
Here, $I$ is the unit matrix, and $B$ is defined as
\begin{eqnarray}
B \bx_{1} & = & 0, \\
B \bx_{2} & = & \gamma x_{1}.
\end{eqnarray}
Note that $B$ is the generator of
$M$($e^{B}\bx_{\alpha}=M\bx_{\alpha}$).
$D$ acts on these basis as
\begin{eqnarray}
D \bx_{1,n} & = & 2n\pi i \bx_{1,n}, \\
D \bx_{2,n} & = & 2n\pi i \bx_{2,n} - \gamma \bx_{1,n}.
\end{eqnarray}
We define real basis as
in ($\ref{realbase}$), then $\Dslash$ is represented in this space
as
\begin{equation}
d_{n} =
\left(
\begin{array}{cccc}
0 & 0 & 0 & 2n\pi \\
0 & 0 & -2n\pi & 0 \\
0 & -2n\pi & \gamma & 0 \\
2n\pi & 0 & 0 & \gamma \\
\end{array}
\right) \;\;\;(n\ge 1),
\end{equation}
\begin{equation}
d_{0} =
\left(
\begin{array}{cc}
0 & 0 \\
0 & \gamma
\end{array}
\right).
\end{equation}
The functional determinant of this part (except zero-mode) is
\begin{eqnarray}
\Det^{'}\Dslash & = & \gamma\prod_{n=1}^{\infty} (2\pi n)^{4}, \\
& \sim & \gamma .
\end{eqnarray}
The integration of the zero-mode must be done
separately. 

If $\Dslash$ is derived from the stationary phase
approximation of the time-independent Hamiltonian system, 
it has at least one zero-mode corresponding to 
time-translation symmetry. In this case $\gamma = -dT/dE$
\cite{cont1}, and the result of the integration of the zero-mode 
is proportional to the period $T$: 
\begin{equation}
K = \frac{T}{\sqrt{2\pi\hbar \frac{1}{i}\frac{dT}{dE}}
\sqrt{|\det (M^{'} - I)|}}.
\end{equation}
Here $M^{'}$ denote the monodromy matrix whose parabolic part is
removed.
The factor $\sqrt{2\pi\hbar \frac{1}{i}\frac{dT}{dE}}$ 
is canceled by the Fourier
transformation (\ref{fourier}), then we obtain the well-known
amplitude factor of the Gutzwiller trace formula:
\begin{equation}
\frac{T}{i\hbar\sqrt{|\det (M^{'} - I)|}}.
\end{equation}
Details of the integration of the zero-mode will be 
discussed in \cite{sugita}.

\begin{figure}
\hskip 3cm
\psfig{file=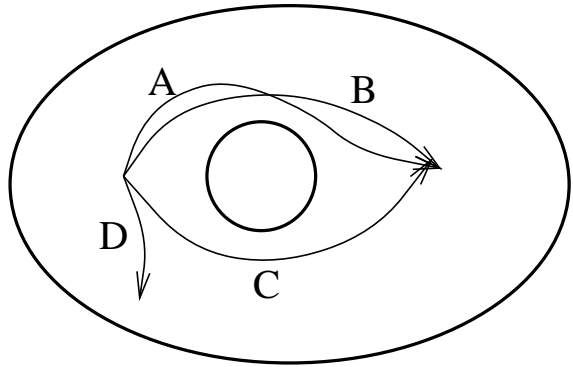,height=0.4\textwidth,angle=-90}
\caption{Curves on the symplectic group. A and B is
equivalent and others are not.}
\label{eq}
\end{figure}

The third step is the calculation of the Maslov index $\mu$.
If $A=0$, the Maslov index is considered to be 0, and
we can see relative changes of the Maslov index
by continuously deforming the operator $\Dslash$.
Thus we can calculate the Maslov index.
The point
is that the deformation of $\Dslash$ corresponds to that of $V(t)$.
$V(t)$ is a curve on the symplectic group, therefore deformation
of $\Dslash$ means deformation of the curve on the symplectic group.
If we continuously deform the curve without changing the
initial point (origin) and the end point (monodromy matrix), 
the Maslov index of the curve doesn't change
because the diagonal elements are determined only by the monodromy
matrix and don't change during the deformation. Therefore we regard
such curves are equivalent (Fig.\ref{eq}). (We refer this equivalence
relation as $\sim$.) 
This means that we can regard two operator $\Dslash (A)$ and 
$\Dslash (A^{'})$ equivalent if $A$ and $A^{'}$ are connected by
a continuous gauge transformation.

However, if two
curves which have the same terminal points can't be shrunk to a point
continuously, two curves doesn't necessarily have the same Maslov
index. Such cases actually happen, because symplectic groups have
non-trivial topology and the fundamental group of them are 
\cite{littlejohn}
\begin{equation}
\pi_{1}({\rm Sp}(2n,R)) = Z.
\end{equation}
In other words, the Maslov index (and the path integral
(\ref{K}))
is the multi-valued
function of the monodromy matrix. The Maslov index is completely
determined by the curve on the symplectic group starting from the
origin (we refer this set of curves as F),
then the Maslov index is uniquely determined on the quotient space
$F/\sim$,
and this is exactly the definition of the universal covering space;
\begin{equation}
F/\sim \;\; = \tilde{{\rm Sp}}(2n,R).
\end{equation}
Therefore the Maslov index is regarded as the function on 
$\tilde{{\rm Sp}}(2n,R)$, instead of ${\rm Sp} (2n,R)$.
This is similar to the construction of Riemann surface of the
complex multi-valued function, for example, $\log z$.

\begin{figure}
\hskip 3cm
\psfig{file=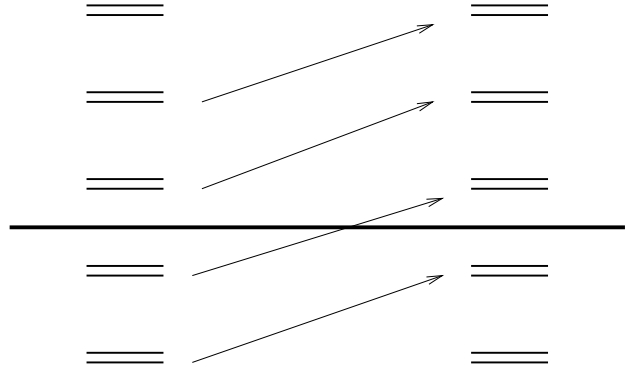,height=0.4\textwidth}
\caption{Flow of diagonal elements of elliptic type when $\alpha$ changes into
$\alpha + 2\pi$. The values of diagonal elements in left side and the
right side are the same, but the Maslov indices differ by 2.}
\label{elliptic}
\end{figure}

A point on $\tilde{{\rm Sp}}(2n,R)$ is specified by two
quantities: a symplectic matrix and a winding number. 
First we investigate the change of the Maslov index which comes
from the winding number. 
For example, if we continuously change 
$\alpha$ to $\alpha + 2\pi$ in elliptic case,
the Maslov index changes by 2. Therefore in this case
the change of the winding number by 1 corresponds to 
the change of the Maslov index by 2 (Fig.\ref{elliptic}). 
This rule can be applied to all cases, 
which can be proven as follows.

\begin{figure}
\hskip 3cm
\psfig{file=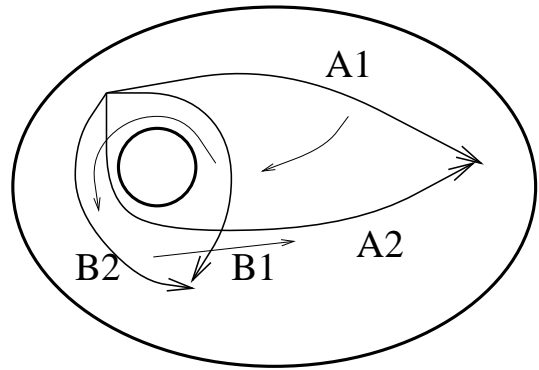,height=0.4\textwidth,angle=-90}
\caption{Changing the windng number. A1$\rightarrow$B1$\rightarrow$B2
$\rightarrow$A2}
\label{wind}
\end{figure}

Let $A1$ and $A2$ be curves on the symplectic group.
We assume that both have the same terminal points and difference
of the winding number between $A1$ and $A2$ is 1.
Then deform $A1$ continuously to $A2$ through the other curves
$B1,B2$ (Fig.\ref{wind}). The difference of the Maslov index 
between $A1$ and $A2$ is the same as that of $B1$ and $B2$,
because the change of the Maslov index during the deformation
$A1\rightarrow B1$ is canceled by that of $B2\rightarrow A2$
if the end-point go through the same path during the two processes.
Therefore the rule that holds for some cases holds for all cases,
and the change of the winding number by 1 corresponds to the change of 
the Maslov index by 2, regardless of the stability.
This means that $(\Det i\Dslash)^{1/2}$ changes sign by the gauge
transformation $U$ that changes the winding number by odd number:
\begin{equation}
\{\Det i\Dslash(A^{U})\}^{1/2} = - \{\Det i\Dslash(A)\}^{1/2}  
\end{equation}
where $A^{U}$ is defined as
\begin{equation}
A^{U} = - U^{-1}\frac{dU}{dt} + U^{-1}AU.
\end{equation}
A similar result is known in  anomaly of SU(2) gauge theory
coupled to Weyl fermions \cite{witten}.

\begin{figure}
\hskip 3cm
\psfig{file=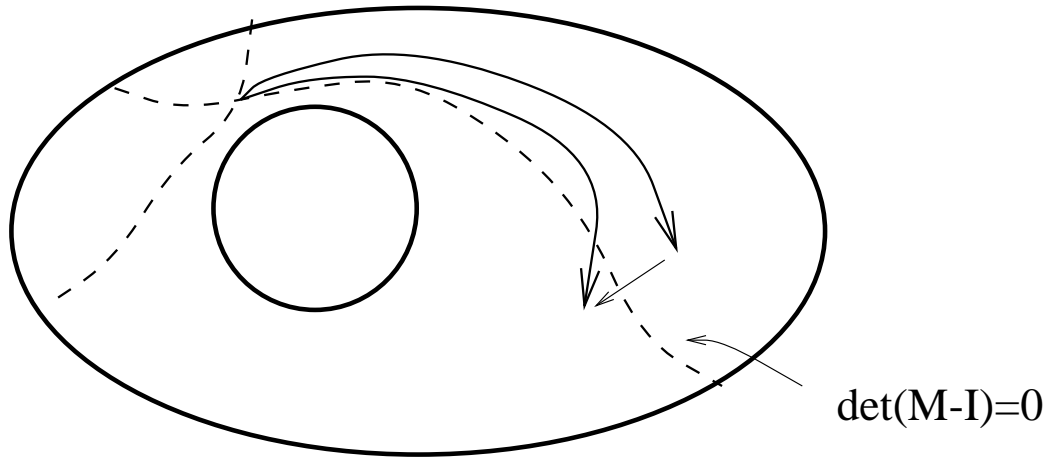,height=0.5\textwidth}
\caption{The Maslov index changes at the point where the end-point
of the curve crosses the region defined by $\det (M-I)=0$.}
\label{Mbif}
\end{figure}

 Now we know the contribution from the winding number to the Maslov
index, the next question is the contribution from the monodromy matrix.
Since the winding number changes the Maslov index by a multiple of
2, the monodromy matrix determines whether the index is even or odd.
The Maslov index changes at the point where $\Dslash$ has zero
eigenvalues, and this means the monodromy matrix has the eigenvalue
1. (This is the condition for bifurcation.)
Therefore the Maslov index changes when end-point of $V$ crosses
the region defined by $\det (M-I)=0$ (Fig.\ref{Mbif}). 

\begin{figure}
\hskip 3cm
\psfig{file=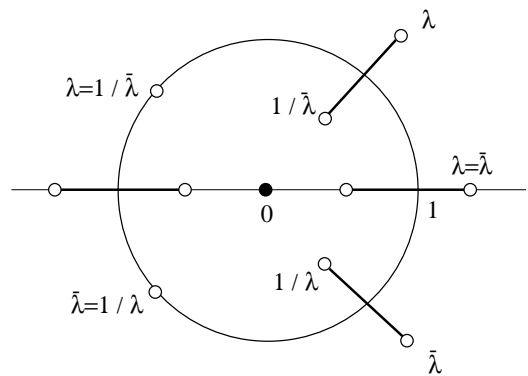,height=0.4\textwidth,angle=-90}
\caption{eigenvalues of a sympectic matrix in the complex plane.}
\label{eigen}
\end{figure}

In Fig.\ref{eigen}, we show eigenvalues of the monodromy matrix 
in the complex plane. The stability changes when eigenvalues collide,
and collisions are classified as follows:
\begin{enumerate}
\item elliptic $\leftrightarrow$ hyperbolic\\
two eigenvalues collide at 1
\item elliptic $\leftrightarrow$ inverse hyperbolic\\
two eigenvalues collide at -1
\item elliptic($\times 2$) $\leftrightarrow$ loxodromic\\
eigenvalues collide simultaneously at conjugate points of the unite circle 
\end{enumerate}
The Maslov index changes when eigenvalues collide at the point 1,
and we can see
from table\ref{table} that the parity of the Maslov index changes
when and only when the stability changes elliptic $\leftrightarrow$
 hyperbolic. 
\footnote{Strictly speaking, the basis (\ref{base})
are not linearly independent when the monodromy matirix has eigenvalue
1. Therefore we must use other basis in this region.}

\begin{figure}
\hskip 0cm
\psfig{file=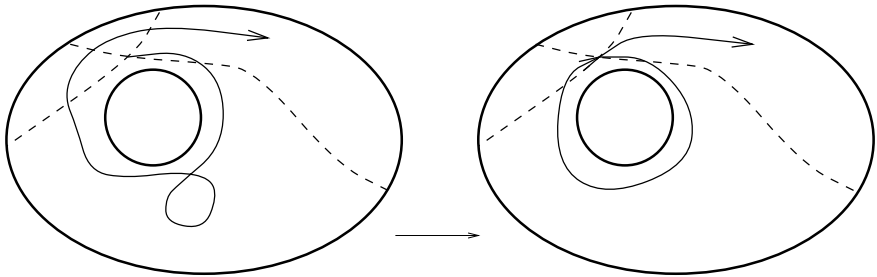,height=0.3\textwidth}
\caption{Deformation of a given curve $C$ into a normal form
$U_{k}C^{'}$. $U_{k}$ starts from the origin, goes around the hole
k times, and returns to the same point. $C^{'}$ doesn't cross 
the region defined by $\det (M-I) = 0$.}
\label{normal}
\end{figure}

A given curve $C$ on the symplectic group can be deformed into a normal form 
$U_{k}C^{'}$ using continuous deformation (Fig.\ref{normal}). 
\footnote{Here we define the product of two curves $A$ and $B$ as
\begin{equation}
AB (t) = \left\{
\begin{array}{cc}
A(2t) & (0 \le t \le 1/2) \\
A(1)B(2t-1) & (1/2 \le t \le 1).
\end{array}
\right. 
\end{equation}
In this definition, $A(1)B(2t-1)$ means the product as matrices.}
$U_{k}$
is the closed curve which have the winding number $k$, and $C^{'}$ is
defined not to cross bifurcation regions. Since the monodromy matrix can be
decomposed into some blocks, $C^{'}$ can also be decomposed into 
these normal forms:
\begin{itemize}
\item elliptic
\begin{equation}
\left(
\begin{array}{cc}
\cos \alpha t & - \sin \alpha t \\
\sin \alpha t & \cos \alpha t
\end{array}
\right) 
\end{equation}
\item hyperbolic
\begin{equation}
\left(
\begin{array}{cc}
e^{\beta t} & 0 \\
0 & e^{-\beta t} 
\end{array}
\right) \\
\end{equation}
\item inverse hyperbolic
\begin{equation}
\left(
\begin{array}{cc}
\cos 2\pi t & - \sin 2\pi t \\
\sin 2\pi t & \cos 2\pi t
\end{array}
\right) \;\;\;(0\le t\le \frac{1}{2})
\end{equation}
\begin{equation}
\left(
\begin{array}{cc}
- e^{2\beta (t-1/2)} & 0 \\
0 & - e^{-2\beta (t-1/2)} 
\end{array}
\right) \;\;\;(\frac{1}{2}\le t\le 1)\\
\end{equation}
\item loxodromic
\begin{equation}
\left(
\begin{array}{cccc}
e^{\beta t}\cos\alpha t & 0 & -e^{\beta t}\sin\alpha t & 0 \\
0 & e^{-\beta t}\cos\alpha t & 0 & -e^{-\beta t}\sin\alpha t \\
e^{\beta t}\sin\alpha t & 0 & e^{\beta t}\cos\alpha t & 0 \\
0 & e^{-\beta t}\sin\alpha t & 0 & e^{-\beta t}\cos\alpha t
\end{array}
\right) \\
\end{equation}
\item parabolic
\begin{equation}
\left(
\begin{array}{cc}
1 &  \gamma t \\
0 & 1
\end{array}
\right)
\end{equation}
\end{itemize}
The elliptic and the inverse hyperbolic type 
have the Maslov index 1,hyperbolic and loxodromic type have 0.
Therefore if the monodromy matrix have p elliptic blocks and
q inverse hyperbolic blocks, the Maslov index of the system is
\begin{equation}
\mu = p + q + 2k
\end{equation} 
If the monodromy matrix have parabolic blocks, we must add 
$\sum_{i}\frac{1}{2}{\rm sgn}\gamma_{i}$ to the Maslov index.
But the factor corresponding to time translation symmetry
is canceled by the Fourier transformation.

We are now able to calculate n-repetition of this orbit.
Since $C^{n}=U_{k}C^{'}U_{k}C^{'}....U_{k}C^{'}$ can be deformed
into  $U_{kn}C^{'n}$, we only have to consider n-th power of the 
normal forms. The elliptic type increases the winding number 
when $n\alpha$ goes over $2\pi$, therefore the Maslov index of this
part is $1+ 2[n\alpha/2\pi]$. The square of the inverse 
hyperbolic type becomes hyperbolic type, and the Maslov index increase
by 1. The cube of it is inverse hyperbolic, the Maslov index increases
by 1, and so on. Therefore the Maslov index of this part is n. 
The hyperbolic type, the loxodromic type, and parabolic type
don't change the Maslov index when we raise them to n-th power.
Thus the Maslov index of the n-repetition of the orbit is
\begin{equation}
\mu_{n} =
\sum_{i=1}^{p}\left(1+2\left[\frac{n\alpha_{i}}{2\pi}\right]\right) +
qn + 2nk,
\end{equation}
where $\alpha_{i}$ denotes the stability angle of i-th elliptic block.
Note that $\mu_{n}\ne n\mu_{1}$ if the orbit has elliptic blocks.

%\section{Conclusion and discussion}
In conclusion,
we have derived the semi-classical trace formula using phase space 
path integral. The point is that quadratic
path integral around the periodic orbit is determined by the
symplectic flow around the orbit, and the flow is regarded as 
a curve on symplectic group. We directly diagonalize the quadratic
path integral, and Maslov indices are defined by the signs of
these diagonal elements. Since they are determined
only by the monodromy matrix, two curves on symplectic group
which have the same
terminal points and can be shrunk to a point is considered to be
equivalent. The set of curves divided by this equivalence relation
become universal covering space of symplectic group, and Maslov 
indices are uniquely determined in this space.  A point 
in the space is specified by a symplectic matrix and 
a winding number. The winding number changes the Maslov index
by 2, and the stability of the orbit determines the parity
of the Maslov index. We also derived the formula for the Maslov
index of n-repetition of a orbit. If the monodromy matrix of the
orbit have elliptic eigenvalues, $\mu_{n}$ is not generally 
equal to $n\mu_{1}$, as pointed out by Brack et al.\cite{brack}

The author would like to thank Professor T.Hatsuda for helpful
discussions and encouragements. He would also like to thank   
Professor M.Sano,
Dr. S.Sugimoto, 
Professor T.Kunihiro, Professor H.Kuratsuji, Professor 
A.G.Magner and the members of
the Nuclear Theory Group at Kyoto University
for valuable discussions.

\begin{table}
\begin{tabular}{|c|c|c|c|c|} \hline
  	 stability & $\lambda$ & diagonal elements 
& $|\Det\Dslash| $ & index\\ \hline
elliptic   &   $e^{\pm i\alpha}$    & 
$\alpha + 2n\pi$& $4\sin^{2}\alpha/2$ & odd\\ \hline
hyperbolic &   $e^{\pm \beta}$     & 
$\pm\beta,\pm\sqrt{\beta^{2}+(2n\pi)^{2}}$ & 
$4\sinh^{2}\beta/2$ & even \\ \hline
inverse hyperbolic & $-e^{\pm \beta}$  & 
$\pm\sqrt{\beta^{2}+(2n+1)^{2}\pi^{2}}$ &
$4\cosh^{2}\beta/2$ & odd \\ \hline
loxodromic &   $e^{\pm i\alpha \pm \beta}$ &
$\pm\sqrt{\beta^{2}+(\alpha +2n\pi)^{2}}$ &
$4(\cosh\beta - \cos\alpha)^{2}$ & even \\ \hline
\end{tabular}
\caption{Stability, eigenvalues of the monodromy matirx, diagonal elements
of $\Dslash$, functional determinant $\Dslash$ of the part, and the
parity of the Maslov index. All diagonal elements are doubly
degenerate except $\pm \beta$ in hyperbolic case. $n$ runs over 
all integers in elliptic and loxodromic cases. In hyperbolic case,
$n\ge 1$, and in inverse hyperbolic case, $n\ge 0$.}
\label{table}
\end{table}

\begin{figure}
\hskip 3cm
\psfig{file=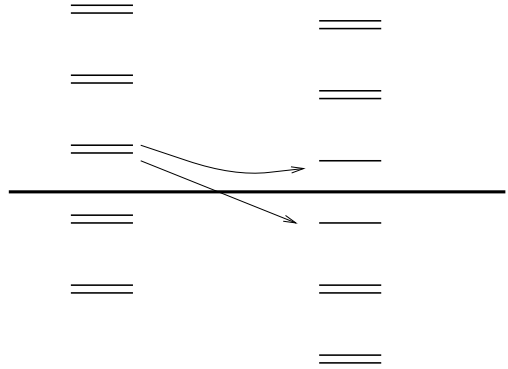,height=0.4\textwidth}
\caption{The change of the sign of the diagonal element of $\Dslash$
when the stabilty changes elliptic to hyperbolic.}
\label{parity}
\end{figure}

\end{document}